\documentstyle[twocolumn,aps]{revtex}

\begin{document}

\title{Is there a symmetry between absorption and amplification in
disordered media?}

\author{Xunya Jiang, Qiming Li, and C.~M.~Soukoulis}

\address{Ames Laboratory and Department of Physics and Astronomy,\\
Iowa State University, Ames, IA 50011\\}

\author{\parbox[t]{5.5in}{\small
Previous studies found a surprising symmetry between absorption and
amplification in disordered one-dimensional systems once the system
size exceeds certain thresholds. We show that this paradoxical result
is an artifact due to the assumption of a finite output in solving the
time-independent wave equation, when in fact both the transmission and
the reflection amplitude are divergent as a result of the large
amplification.
More sophisticated approaches, such as time-dependent equations
including field and matter coupling have to be utilized to correctly
describe the wave propagation of systems with large sizes. 
 }}

\maketitle

\newpage

Recent observations \cite{expt} of laser-like emission from dye
solution emerged with TiO$_2$ nanoparticles have stimulated intensive
theoretical
efforts\cite{john2,pr,zhang,gupta,sen,been,li,joshi,kim,jiang,pradhan}
to investigate the properties of disordered media which are optically
active. From considerations of enhanced optical paths through multiple
scattering, random systems are expected to possess a reduced gain
threshold for lasing\cite{letok,zyu}. Correspondingly, one would expect
the transmission in disordered systems to be enhanced with gain. The
longer the system, the larger the enhancement. Surprisingly, numerical
calculations\cite{zhang} based on time-independent wave equations
showed that for large systems the wave propagation is still attenuated,
indicating a
localization\cite{anderson,john} of waves even with gain. The rate of
the exponential attenuation is the same as if the system was absorbing.
Such a symmetry between absorption and amplification was subsequently
shown\cite{dual} to hold for the time-independent wave equation.

Intuitively one would expect that the presence of amplification should
facilitate wave propagation, not suppress it, even in disordered
systems. The peculiar result of reduced transmission in gain media is
also absent when wave propagation in
disordered media
is described by time dependent diffusion equations which always predict
an increased output and a gain threshold above which the system become
unstable\cite{letok,zyu,burkov}. The diffusion equation neglects the
phase coherence of the wave and is adequate only when the wavelength is
much smaller than the mean free path. Thus it was
proposed\cite{zhang,joshi} that
the paradoxical phenomenon may indicate enhanced localization due to
the amplification of coherent backscattering. However, amplification of
backscattering does not necessarily imply a reduction in transmission
since no conservation of the total photon flux
is required in gain media. To fully understand the origin of this
non-intuitive result, we will examine the validity of the solutions
derived from the time-independent wave equation which has been commonly
employed
in describing the wave propagation in active media.

Linearized time-independent wave equations with a complex dielectric
constant
have been successfully utilized to find lasing modes by locating the
poles\cite{yuriv} in the complex frequency plane and to investigate the
spontaneous emission noise below the lasing threshold in distributed
feedback semiconductor lasers\cite{henry,makino}. Such equations are
known to be inadequate\cite{yuriv,henry} to describe the actual lasing
phenomena due to their simplicity in dealing with the interactions
between radiation and matter. However, it is generally believed that
the time-independent equation should suffice to describe
the wave propagation in amplifying media, before any oscillations
occur. We will unambiguously show that the so-called
symmetry\cite{dual} between amplification and absorption is an artifact
due to the unphysical assumption of a finite output in solving the wave
equations. We show that for each system, there is a
frequency-dependent
gain threshold above which both the total transmission and the total
reflection become divergent. Solving the time-independent wave
equations by incorrectly assuming a fixed output leads to unphysical
solutions that does not correspond to the true behavior of the system.

To demonstrate our point, we take the simplest example of a uniform
active media sandwiched between two mirrors as feedback (see the insert
in Fig. 1), the classical Febry-Perot setup. Wave propagation within
the active media is simply described with the following
phenomenological wave equation, 

\begin{equation}
\frac{d^2 E(z)}{dz^2} + \frac{\omega^2}{c^2} \varepsilon (z) E(z) = 0 ,
\end{equation}
where $E(z)$ is the electric field and the dielectric constant
$\varepsilon(z) = \varepsilon^\prime(z) - i\varepsilon^{''}(z)$ with
the imaginary
part signifying amplification ($\varepsilon^{''} >0 $) or absorption
($\varepsilon^{''}< 0$ ). We point out that in electromagnetic theory,
the imaginary part of the dielectric
constant is proportional to the conductivity of the material and thus
cannot be negative. The negative dielectric constant is strictly
speaking only an effective way to introduce coherent amplification
\cite{lamb}. Complex potentials known as optical potential have long
been employed in nuclear physics to describe nuclei scattering
processes.

The transmission and the reflection amplitude can be easily obtained by
solving Eq.(1) to yield: 
\begin{equation} \label{Fa}
t = \frac{ t_1t_2e^{ikL} } {1-r_1r_2e^{2ikL}} 
\end{equation}
where $t_1=2k/(k+k_0)$, $t_2=2k_0/(k+k_0)$, and
$r_1=r_2=(k-k_0)/(k+k_0) $ are the transmission and reflection
coefficients at the left and right two interfaces, respectively.
$k_0=\sqrt{\varepsilon_0}\frac{2\pi}{\lambda}$ and
$k=k'-ik''=\sqrt{\varepsilon'-i\varepsilon''}\frac{2\pi}{\lambda}$ are
the wavevectors outside and inside the system, $L$ is the distance
between the mirrors
(system size). The oscillation condition for lasing is correctly given
by $1-r_1r_2e^{2ikL}=0$ at which both the transmission and reflection
coefficient diverge.

However, Eq.(2) also predicts the exponential decrease of the
transmission coefficient
for large system sizes. In fact the term in the numerator with the
$exp(ikL)$ increases exponentially as
the length of the system $L$ increases because of the gain, but the
term in the denominator with the $exp(2ikL)$ increases even faster,
making the transmission coefficient decay asymptotically as $|t| \sim
|exp(-ikL)|=exp(-k''L)$. This is clearly shown in Fig.1, where we plot
the $ln(T)$ versus $L$ for a one-layer system. Notice that for large
$L$, $ln(T)$ decreases as $L$ increases despite the fact that we have
gain at here. Gain effectively
becomes loss at large lengths! Remember the system is homogeneous thus
disorder is definitely not responsible for this strange behavior. Thus
the inhibition
of wave propagation for large systems is clearly not a result of
amplification in
backscattering.

A clear picture of what is going on can be obtained from the path
integral method \cite{yuriv}. In such an approach, the total
transmission coefficient
can be obtained by adding the paths of the successive reflected and
transmitted rays. In doing so, we obtain that 

\begin{equation}
t= t_1t_2e^{ikL}
[1+(r_1r_2e^{2ikL})+(r_1 r_2e^{2ikL})^2 + \cdots] \label{path}
\end{equation}
where the first term represents the direct transmission of the incoming
wave, and the
second term represents the wave which was reflected first by $r_2$ at
right interface and then by $r_1$ at left interface and subsequently
transmitted through. More terms from sequences of multiple
transmissions
and reflections from the two mirrors follow. It is clear Eq.(3)
represents an infinite series whose sum reproduces Eq.~(2), provided
that the following condition is met,

\begin{eqnarray}
|{r_1}{r_2}{e^{2ik'L}e^{2k''L}}|<1 \label{over}. 
\end{eqnarray}

When the condition given by Eq.(\ref{over}) is violated, such as when
the system size or the gain is large,
the physical output represented by the sum diverges, even away from the
oscillation
pole. 

We hope to note at here that the sum of right side of Eq.(3) include the 
phases of all paths, so that it  includes all interference of 
scattering waves. We also note that the results of Eq.(3) are consistent 
with time-dependent theory because different terms of Eq.(3) can explain 
as   
outputs at different time from a same incidence or the output at same time 
from a series incidence at different time.   
By this way, the out-put of a system with gain much larger than the 
threshold will increase exponentially versus time after  a 
plane wave incidence, 
 but time-independent theory suppose a small output as if the system is of 
absorbing material. 
Base on time-dependent Maxwell equations, a 
well-developed  FDTD (finite-difference time-domain) method can help 
us  to see  the obvious different behavior of transmission of two kind of 
systems versus time. We choose a plane wave with $\lambda=800$nm as 
the incidence from left to the  setup of Fig.1 with $ L = 
4300 $nm. The dielectric constant is taken to be $\varepsilon_0=1$ 
outside and $Re(\varepsilon)= \varepsilon'=9$ inside. The threshold of 
this system is $\varepsilon''=0.12$ from Eq.(4). Our numerical results 
show that, after a 
short instantaneous state,  the output of system will get to a stable value 
which is the transmission of the system if the 
system with under-threshold gain or with absorption. Unless the gain or 
the length is over the threshold, the results of time-independent 
theory are same as the time-dependent theory.  
But if the system with gain is above the threshold, the output 
of system will increase exponentially versus time  as 
predicted by Eq.(3). 
In Fig.~2, we 
plot the logarithm of amplitudes of out-put at right 
side of system versus time for different gain, we can see the numerical 
results  exactly same as we predicted.
We also examined the slope of lines with the  gain larger the threshold 
in Fig.~2, 
and found the slope is exactly same as we predict in Eq.~(3), equal to 
$log_{10}|{r_1}{r_2}e^{2k''L}|$. 

The divergence of the transmission above threshold even away from
the oscillation pole is the key in understanding the failure of Eq.(2),
which is conventionally derived from boundary condition by implicitly
assuming that the output is
finite.
Normally the physical boundary condition is satisfied when
$t_1t_2e^{ikL}+r_1r_2e^{2ikL}t =t$, resulting in Eq. (2). Obviously
this condition lost its meaning when t is infinite. As a result, Eq.
(2) becomes unphysical. For a lossy system, the condition given by
Eq.(\ref{over}) is always satisfied. Thus the divergence is a new
phenomenon occurring only to systems with gain. The breakdown of the
time-independent
wave equation signals the large fluctuations of the transmission in
time and calls for
more sophisticated theories that can take into account the interaction
between radiation and
matter to correctly describe the response of the system.

To show that the exponentially attenuated results of time-independent theory
with gain is not from the backscattering effect, but from a theoretical
mistake, we also do the same numerical experiment to random system with
gain. The system is made of 50 cells, each cell contains two kind layers 
with dielectric constant 
$\varepsilon_1=1$ and $\varepsilon_2=4-i\varepsilon''$.  To introduce 
disorder, we choose the width of first layer of the $n$th cell to be 
random variable ${a}_n={a}_0(1+W\gamma)$,where ${a}_0=95$ nm, $W$ 
describes the
strength of randomness which is 0.8 in our model and $\gamma$ is a random 
number between $(-0.5,  0.5)$. The width of second layer in $n$th cell is  
${b_n}=215nm-{a_n}$. The wavelength of incident wave is $\lambda=1200$nm.
In Fig.3, we 
plot the logarithm of output at right side of system versus time
with plane wave incidence  for different gain.
We also get the same 
results as what we predicted, the exponential 
increase of the output versus time when the gain is over the threshold
value, which are
showed in Fig.3.  The time-dependent results 
also show that the 
threshold of random system is quite small as previously predicted and
showed experimentally \cite{expt}.
Actually all time-dependent theories about the system with the gain or
the length over the threshold, such as Letokohov diffusion theory
\cite{letok}, Lamb theory \cite{lamb}, give the divergent output of the
system. The physical meaning of the divergent output is that
the rate of generating photons by the induced radiation is larger than the
escaping rate of photons from the interfaces of the system. Now the field 
in the system will become stronger and stronger even if the frequency 
of the
wave is not resonant frequency(at the resonant frequency, the system 
become a laser).  

 For multilayer systems, to see 
whether the results of transmission or
reflection coefficients of the time-independent theory is physical, a
simple approach is to check every layer of the
system by the following method. When we examine one certain layer, such
as m-th layer, in the multilayer system, we can assume that the left
part of system forms an effective interface of the layer with
transmission and reflection
coefficients $t_l$ and $r_l$, and the right part forms the other
effective interface with $t_r$ and $r_r$. Here we assume both right and
left subsystems are below threshold.
The convergence condition for every layer of the system is then given
by Eq.(2), with ($t_1$, $r_1$),
and ($t_2$, $r_2$) replaced by ($t_l$,$r_l$) and ($t_r$, $r_r$),
respectively. For every layer, Eq.(2) gives us a line. The cusps are
formed by the lowest limit out of all lines. Fig.~4
shows the results of such a calculation for a 40 layer system, which is
formed by two kind of layers with widths $L_1=95 nm$, $L_2=120 nm$ and
dielectric constants $\varepsilon_1'=1$, $\varepsilon_2'=2$
respectively . 
In reality, the curve only gives an indication of the magnitude of the
gain
above which the results from time-independent equation becomes suspect.

In solving the time-independent equations, it is difficult to know
exactly when the solutions
break down. One certainly cannot tell from the expression of the total
transmission
and the total reflection coefficient which are well behaved except
exactly at the oscillation
pole. However, a rough indication is that when the gain for a fixed
system approaches the
lasing threshold for nearby poles or when the system sizes exceed the
threshold length,
the solutions should not be trusted. However, the threshold condition
is still given
by the poles in the complex plane. A study of the distribution of these
poles has been
carried out through carefully locating the pole position by
continuously tuning the gain up at a fixed frequency until the
transmission coefficient diverges\cite{li}.

The implication of the analysis above is that the calculation of
transmission and reflection coefficient with the traditional method
become suspect once the gain or system size reaches a certain value.
Care has to be taken to ensure that the system is not above threshold
and the solution is physical. This remark, unfortunately, also pertains
to the application of the powerful $invariant$ $embedding$ method
\cite{pr,been,joshi,kim}. in which the finite reflection and
transmission are assumed implicitly.

In summary, this note aims to bring attention to some peculiar aspects
of the time-independent wave equation when the gain is above certain
threshold for a fixed length, or equivalently, when the length of the
system exceeds a certain value for a fixed gain. Thus some of the
conclusions on the statistical
properties of the reflection and the transmission coefficient in media
with gain become suspicious at large
system sizes. The time-independent equation is inadequate to describe
the amplification of light under these conditions. Nevertheless, the
simplicity of the time-independent equation can be used effectively to
locate resonant conditions, even in disordered systems. A complete
treatment of the wave propagation in gain media may require the
construction of the time-dependent solution out of the continuous and 
discrete solutions of the time-independent equation \cite{economou}. 
Unlike for the Hermitian system, the completeness of these solutions
could not be proved easily \cite{opticalpotential} when the potential
is complex.

Ames Laboratory is operated for the U.S. Department of Energy by Iowa
State University under Contract No. W-7405-Eng-82. This work was
supported by the director for Energy Research, Office of Basic Energy
Sciences.

\begin{figure}
\caption{
The logarithm of the transmission coefficient, $ln(T)$, versus the
length $L$ of a simple Febry-Perot setup. The dielectric constant of the
Febry-Perot device is
$\varepsilon=\varepsilon'-i\varepsilon''=2-i0.01$. The dielectric
constant of outside medium is $\varepsilon=1$ without gain.\\ 
In the insert a schematic representation for wave transmission in a
simple Febry-Perot setup is shown.}
\end{figure}

\begin{figure}
\caption{ The logarithm of output at right side of system versus time of 
a Febry-Perot setup with a plane wave ($\lambda=800$ nm)incidence. 
The length of system is the $L=4300$ nm and the dielectric
constants of inside and outside materials are
$\varepsilon_1=9-i\varepsilon''$ and $\varepsilon_1=1$ respectively. The 
time unit $t_0$ is the time for the wave to travel back and forth through 
the system. The critical gain of the system is $\varepsilon''_c=0.12$. } 
\end{figure}

\begin{figure}
\caption{ The logarithm of output at right side of system versus time of
a random system with a plane wave($\lambda=1200$ nm) incidence. The 
system is made of 50 
cell and each cell contain two kind layers with dielectric constant 
$\varepsilon_1=1$ and $\varepsilon_2=4-i\varepsilon''$ respectively. In 
$n$th cell, the 
width of first layer  is  $a_n={a}_0(1+W\gamma)$, where $a_0=95$ nm,
$W=0.8$ and $\gamma$ is a random number between $(-0.5,  0.5)$, the width 
of second layer is $b_n=215 -a_n$ nm. The time unit $t_0$ is the time for 
the wave to travel back and forth through the system with phase speed.} 
\end{figure}

\begin{figure}
\caption{
The negative value of the imaginary part of the dielectric constant
$\varepsilon''$ versus the incident wavelength $\lambda$ for a 40-layer
system with two kind of layers of width $L_1=95nm$ and $L_2=120nm$, and
with dielectric constant $\varepsilon'_1=1$ and $\varepsilon'_1=2$
respectively. The solid line shows the maximum value of $\varepsilon''$
which the solution of the time-independent equations can be trusted.}
\end{figure}


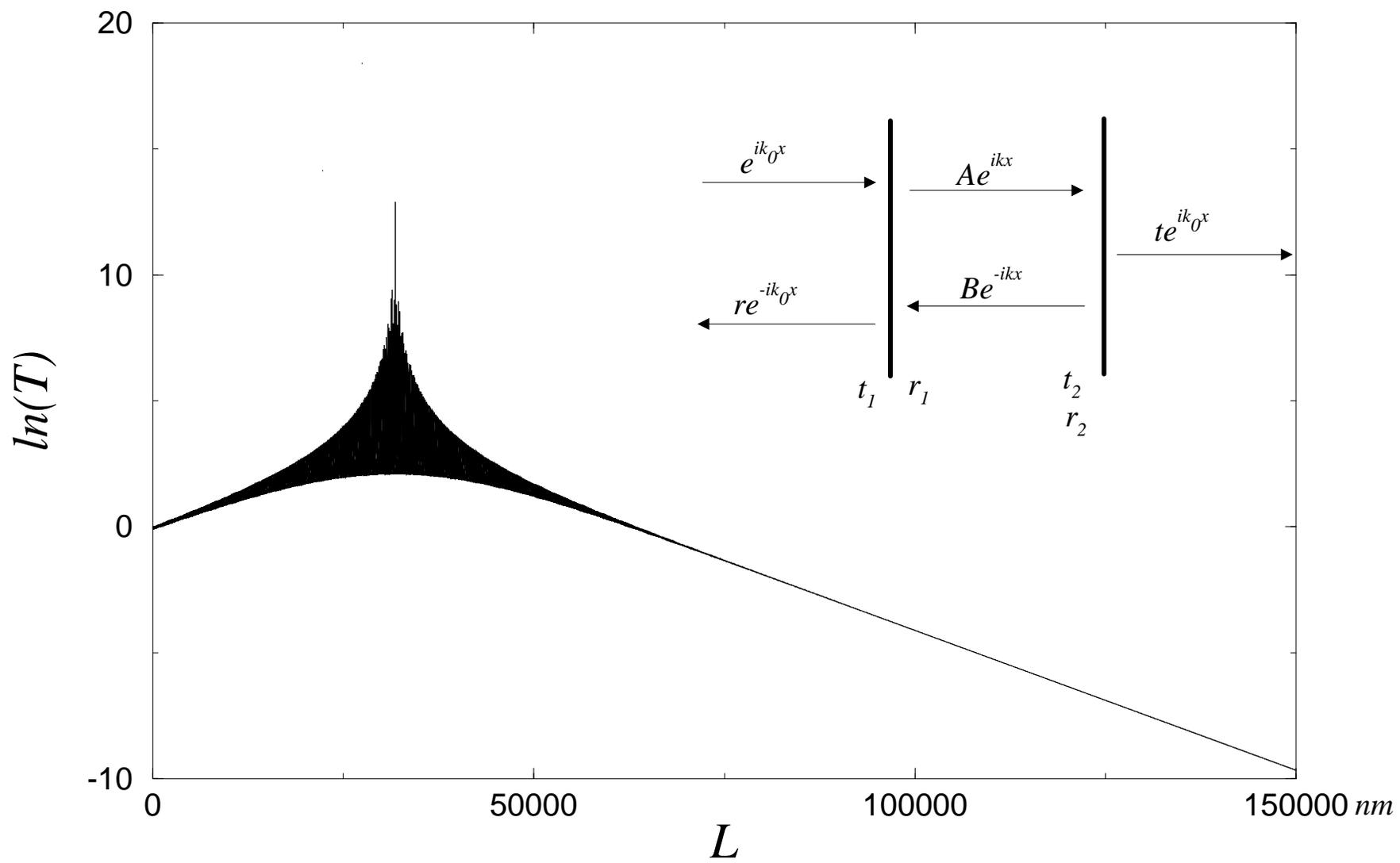
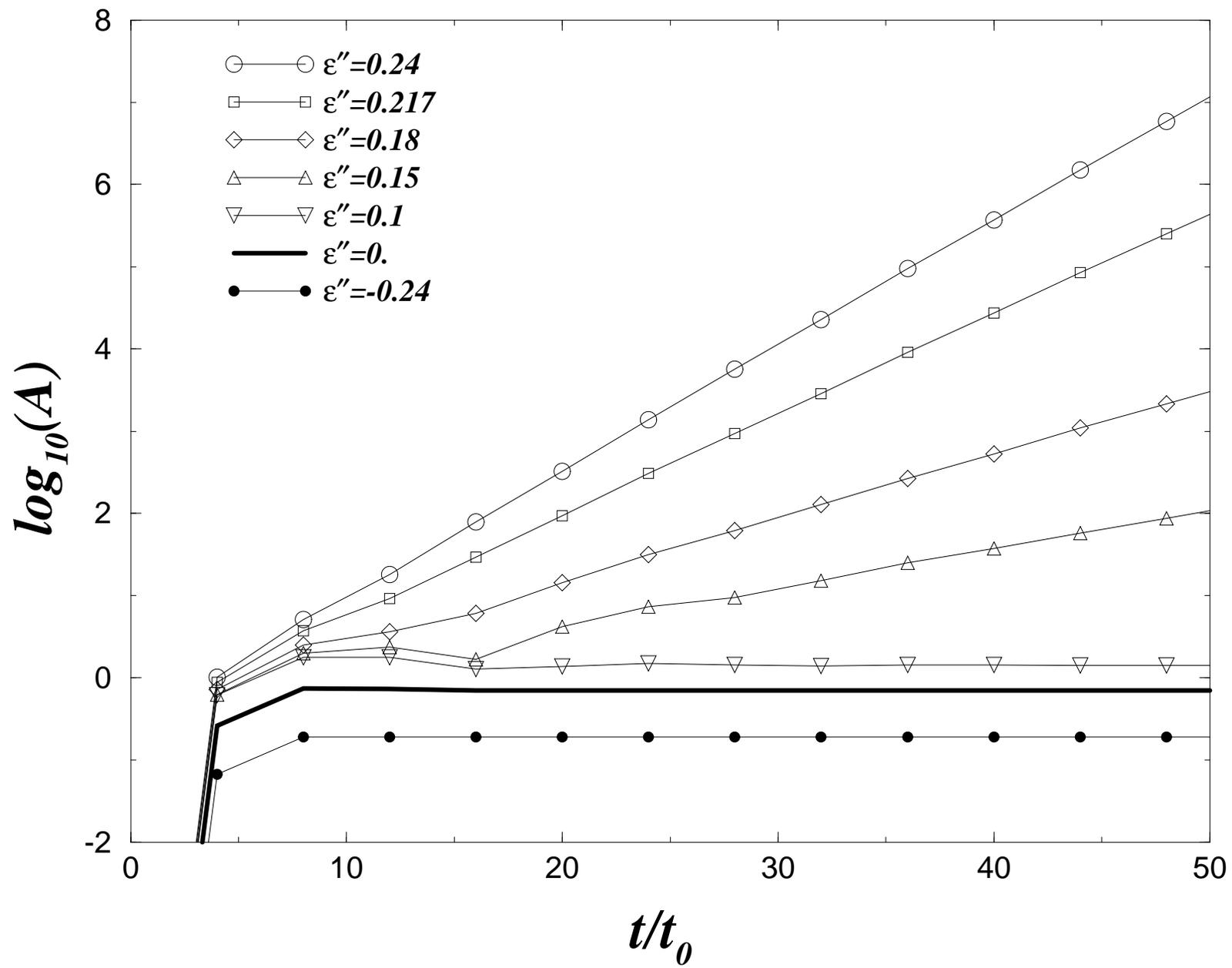
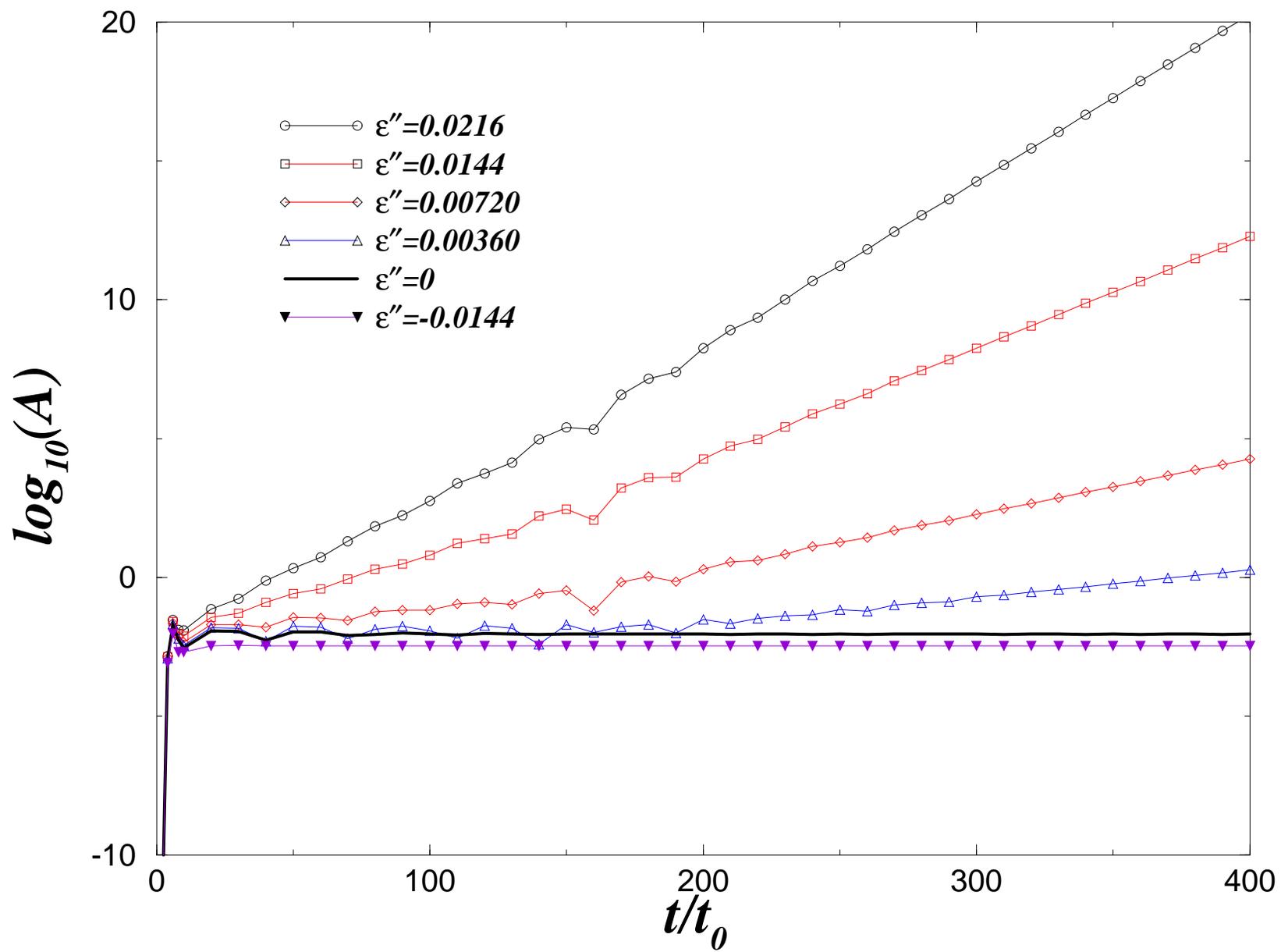
\end{document}